\documentstyle[prl,aps,multicol,epsf]{revtex}

\begin{document}
\draft

\title{Tunneling via individual electronic states in ferromagnetic
nanoparticles}

\author{S.~Gu\'eron, Mandar M.~Deshmukh, E.~B.~Myers, and D.~C.~Ralph}
\address{Laboratory of Atomic and Solid State Physics, Cornell University,
Ithaca, NY 14853}
\date{\today}
\maketitle

\begin{abstract}
We measure electron tunneling via discrete energy levels in ferromagnetic
cobalt particles less than 4~nm in diameter, using non-magnetic electrodes.
Due to magnetic anisotropy, the energy of each tunneling resonance shifts
as an applied magnetic field rotates the particle's magnetic
moment.  
We see both spin-increasing and decreasing tunneling transitions,
but we do not observe the spin degeneracy at small magnetic fields seen
previously in non-magnetic materials.
The tunneling spectrum is denser
than predicted for independent electrons, possibly due to spin-wave
excitations.
\end{abstract}


\begin{multicols} {2}
\narrowtext
The forces that determine the electronic properties of semiconductor
quantum dots or metal nanoparticles
can be investigated in a particularly direct, fundamental way
through tunneling measurements of the discrete ``electrons-in-a-box"
spectrum of energy levels.
This technique has been used to study the
quantum-Hall-effect regime  \cite{Ray},
superconducting pairing in aluminum particles \cite{Black,Braun}, and
effects of more generic electron-electron interactions
\cite{Tarucha,Stewart,Agam,Drago}.
In this Letter we turn
to a ferromagnetic material (cobalt), with the aim of probing the ways
in which strong
exchange interactions and magnetic anisotropy affect the discrete
electron spectrum, as well as investigating spin-polarized tunneling via
single quantum states.  Our work may be viewed as an extension to
smaller size and/or lower temperature of previous experiments
employing micron-size ferromagnetic islands
\cite{Ono} and nm-scale cobalt particles \cite{Schelp}.
We find a number of
phenomena different from past studies of non-magnetic
nanoparticles.  
We observe that the energy levels are coupled to the direction of the particle's total magnetic moment, such that the levels shift as the moment is reoriented.
As might be expected, 
there is no degeneracy in Co between spin-up and 
spin-down tunneling levels
near zero magnetic field, $H$.  
The energy spacing of the
resonances is smaller than expected in an
independent-electron model, suggesting the importance of low-energy
many-body spin excitations.

Our samples consist of Co particles connected to Al leads via
tunnel barriers.  The fabrication is similar to 
previous work on Al particles \cite{Ralph}, and a schematic sample
cross-section is shown in Fig.~1(a).  The top Al electrode is deposited
first so as to fill a bowl etched through a Si$_3$N$_4$ membrane (hole
radius $<5$~nm), and then a tunnel barrier is formed by oxidizing
the Al. The Co nanoparticles are obtained by evaporation at
room temperature of a Co layer 0.5~nm~thick.
Surface tension causes the Co
to form electrically separate particles.  Scanning transmission
electron microscope (STEM) images of test samples in
which 0.5~nm of Co is deposited on oxidized aluminum indicate Co particles
with diameters 1-4~nm, with center-to-center spacing 2-5~nm (Fig.~1(b)).  
\linebreak
\begin{figure}
\vspace{-1.2 cm}
\begin{center}
\leavevmode
\epsfxsize=8 cm
\epsfbox{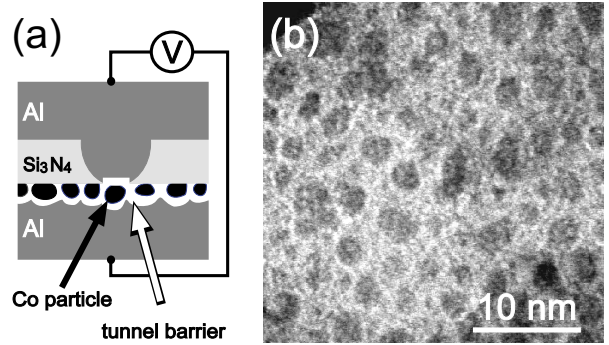}
\end{center}
\vspace{-5.7 cm}
\caption{
\label{figure1}
(a) Cross-sectional device schematic.
(b) Annular dark-field plan-view STEM image of Co particles.}
\end{figure}
\noindent
The crystal structure was not determined.  The second 
tunnel barrier is formed on the Co particles either by
depositing a 0.8 nm-thick layer of Al at 77~K, which we oxidize in
50~mTorr of O$_2$ for 3~min at room temperature (sample 1), or by directly
depositing
1.1 to 1.5 nm of Al$_2$O$_3$ (samples 2,3). Finally a
thick layer of
Al is deposited to make the second electrode.  We select devices for which
the current-voltage curve at 4.2 K shows Coulomb staircase structure (not
shown), indicating tunneling via nanoparticles \cite{many}.

	Figure~2 shows the tunneling spectra at the onset of conduction for the
first Coulomb threshold, for three samples.  The spectra consist of
well-resolved peaks due to tunneling via discrete electronic levels within
each particle, qualitatively similar to previous measurements in Al and Au
\cite{Ralph,Drago}.  The values on the energy axis are determined by
dividing the
voltage by (C$_1+$C$_2)/$C$_2$, to correct for capacitive division
of the bias, where C$_1$ and C$_2$ are the capacitances of the particle to
the two electrodes \cite{Ralph}.  
We can
determine this ratio to within 20\%  by fitting the
temperature-dependent broadening of peaks \cite{Chuck}, or in one sample
having no voltage-dependent charge shifts (sample 2), we can achieve 1\%
accuracy by comparing the voltages required for tunneling via the same
electronic states at positive and negative bias \cite{Ralph}.  The peak
spacing for all three samples is much less than the Coulomb 
\linebreak
\begin{figure}
\vspace{-1.2 cm}
\begin{center}
\leavevmode
\epsfxsize=7 cm
\epsfbox{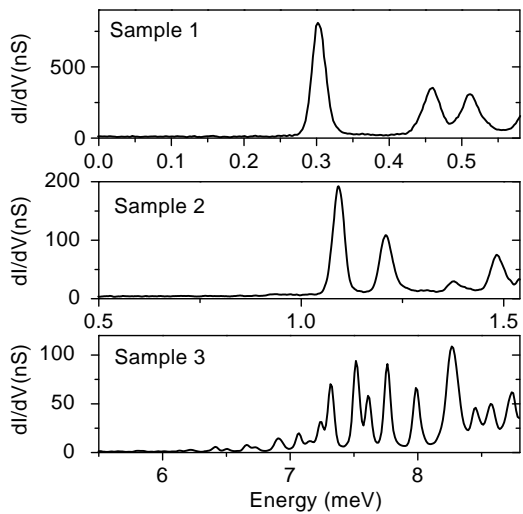}
\end{center}
\vspace{-2.5 cm}
\caption{
\label{figure2}
Tunneling spectra of 3 different samples at
T=20 mK and $\mu_0H=1$~T .  $H$ is parallel to the Si$_3$N$_4$ membrane.
The energy is obtained by dividing $V$ by
(C$_1$+C$_2$)/C$_2=$ 1.2, 2.17 and 2.5, respectively,
for the three samples.}
\end{figure}
\noindent
charging energy
($> 30$ meV, based on particle size), so 
that all peaks in each spectrum
correspond to tunneling 
via states with the same number of electrons,
either one more or one less than the initial state.

Unlike non-magnetic particles where the energy levels have a simple
linear dependence on $H$ due to the spin
Zeeman energy \cite{Ralph}, 
the levels in Co 
have a strong, non-linear dependence on $H$ for small $H$.  Figure~3(a)
shows the energy of the first three tunneling resonances of sample 1, as
$\mu_0H$
was swept about a hysteresis loop from $-0.45$~T to $0.45$~T and back
again.  Starting at $-0.45$~T (thick lines), the tunneling energies shift
in a continuous
manner as the field is ramped to zero and the magnetic moment vector
$\vec{m}$ of the particle relaxes toward its easy direction.  As the field
is ramped further, we observe a sudden jump in all three transition
energies,  at $\mu_0H_{\rm{sw}}=0.23$~T.  We interpret this jump as due to
the reversal of $\vec{m}$ in the single-domain Co
nanoparticle \cite{chemical}.
The energy
shifts with $H$ are hysteretic with respect to the
direction of the field
sweep, with the expected symmetry around $H\!=\!0$, and the scans are
repeatable over days.  The value of $H_{\rm{sw}}$ is comparable to
SQUID
measurements of $25\pm5$ nm diameter Co particles \cite{Wernsdorfer}, and corresponds to an anisotropy energy density of order $K$=10$^5$ J/m$^3$.

These curves indicate a significant coupling between the level energies and
the orientation of the magnetic moment of the nanoparticle.  Let us 
consider the simplest model for this anisotropy, so as to see what 
features of the data may be explained simply, and what features may
require deeper understanding.
We call the
operator for the total electronic spin with
$N$ electrons $\hbar\vec{S}(N)$, and
we assume that the anisotropy and
Zeeman energies are sufficiently weak relative to the exchange splitting
between different spin multiplets that the magnitude $S$
in the ground state remains
\linebreak
\begin{figure}
\vspace{-1.14 cm}
\begin{center}
\leavevmode
\epsfxsize=7.5 cm
\epsfbox{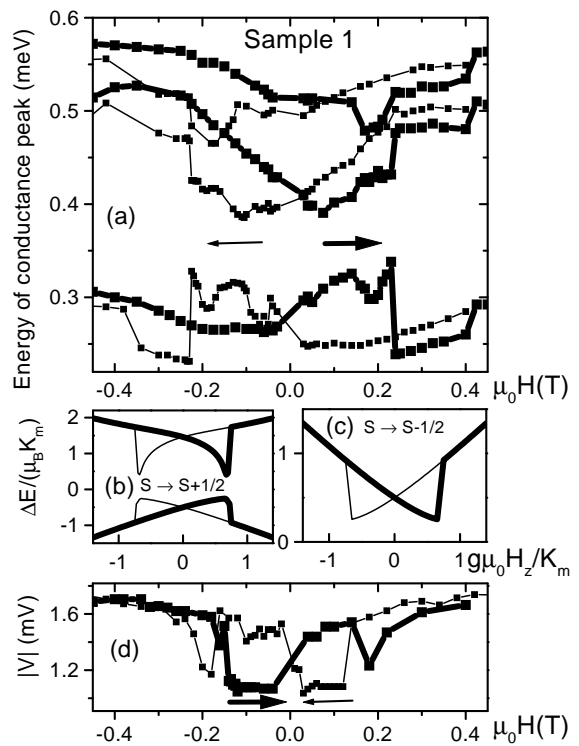}
\end{center}
\vspace{-0.0 cm}
\caption{
\label{figure3}
(a)  Hysteresis curves showing the dependence of tunneling energies on $H$
for sample~1, at T=20~mK.  (b,c) Lowest-energy  transitions
calculated using the Hamiltonian discussed in text, for S=50 and $H$
oriented 45$^{\circ}$ from the easy axis, for the case where $S$
increases during tunneling (b) and decreases (c).  The qualitative features
are independent of the value of $S$.
(d) Voltage threshold for tunneling in one sample which exhibits
anti-hysteretic behavior.}
\end{figure}
\noindent
constant as $H$ is varied, so that we can perform our 
calculation in the
space of this one spin multiplet.  (In 
a classical language this
corresponds to assuming that the ground state magnetic moment $\vec{m}(N)$
simply rotates as a function of $H$.)  Including the Zeeman energy and the
simplest model of easy-axis anisotropy in the $\hat{z}$ direction
\cite{Cullity}, the $H$-dependent terms in the Hamiltonian within the ground-state multiplet for
$N$ electrons can be written
\begin{equation}
{\cal H}=-g_{\rm eff}\mu_{B}\mu_0\vec{H}\cdot\vec{S}
-K_{m}\mu_BS_z^2/\sqrt{S(S+1)}
\end{equation}
where $K_{m}$ is an anisotropy energy per unit $|\vec{m}|$.  
($K_m\equiv$$K$(volume)$/|\vec{m}|$.) 
We have omitted from Eq.~1 the charging energy,
which we assume to be independent of $H$.  For given total
spin values in the $N$ and $N\pm1$-electron states, we diagonalize the
Hamiltonian numerically to find the energy levels, and then calculate the
allowed tunneling transition energies as
$\mathit{E(N\!\pm\!1,H)-\!E(N,H)}$.  We allow the moment
vector of the nanoparticle to undergo reversal at the classical
switching field, which depends on the angle between $\vec{H}$ and the easy axis.  Representative results for the lowest-energy
tunneling transitions are shown in Fig.~3(b) for the case that $S$
increases during tunneling, and in Fig.~3(c) for $S$ decreasing.
The model successfully reproduces the hysteretic energy maxima near zero
field observed in the lowest energy transition (with $S$
increasing), as well as the existence and the sign of the abrupt switching
to lower energy at $H_{\rm{sw}}$.  If we define $S_H$ as the
component of the total electron spin in the direction of $\vec{H}$,
in general we find that tunneling
transitions in
which $|\!\langle$$S_H\rangle\!|$ 
increases give maxima near $H\!=\!0$ and transitions
decreasing $|\!\langle$$S_H\rangle\!|$ give minima.
We can identify both
types of behavior in Fig.~3(a).  We have also solved the classical analogue
of the model, which  gives similar results for the ground-state to
ground-state transitions.

The measured tunneling energies often have small-scale non-monotonic
variations as a function of
$H$ that are not present in our minimal model.
A likely cause
is magnetic interactions between nearby Co particles.  The most
dramatic example we have observed is shown in Fig.~3(d), where we plot the
$H$-dependence of the threshold voltage for tunneling via a Co nanoparticle
too large for discrete resonances to be observed.  We see anti-hysteresis
-- magnetization reversal occurs before $H$ changes sign.
This can
be explained by the influence of a dipolar magnetic field oriented opposite
to the applied $H$, produced by a second magnetic nanoparticle adjacent to
the one through which electron tunneling occurs.  The reversed field from
the second particle can shift the hysteresis curve of the first so that its
value of $H_{\rm{sw}}$ can be negative, while the non-monotonic shifts at
large positive $H$ ($0.2$ T) are understood as the magnetization reversal
of the
second particle.  
A dipole field 5~nm from the center of a 3-nm-diameter Co
particle is of order 0.1~T, so that Zeeman interactions alone are
not strong enough to explain the level shifts that we observe.

A second departure of the data from the minimal model is that 
in some cases the model fails to describe the
combined low and high $H$ variations of the transition
energies (Fig.~4). At large $H$, the simple model predicts a linear
extension of the low field curves, with the tunneling energies
Zeeman-shifting as $\pm$$g_{\rm{eff}}\mu_B\mu_0H/2$.  Figure 4(a) shows
that at large $H$ the ground state transition of sample 1 moves to higher
energies with increasing $H$, indicative of a tunneling transition in which
$|\!\langle$$S_H\rangle\!|$ decreases, whereas the 
low-field behavior indicates a
$|\!\langle$$S_H\rangle\!|$-increasing transition. Similarly, the ground state
transition of
sample 2 (Fig 4(b)) is $|\!\langle$$S_H\rangle\!|$-decreasing 
at low field (dip to
lower energy)
but $|\!\langle$$S_H\rangle\!|$-increasing at high field
(energy shift to lower energies).
We see two possible explanations.  Either ({\it i})
the true form of the anisotropy is more complicated
than assumed in Eq.~(1), such that $K_m$ fluctuates to
have different values for
different spin multiplets, or
({\it ii}) contrary to our model's other initial assumption,
the total spins of the ground-states for $N$ and $N\pm1$ electrons are {\em
not} independent of $H$, so that $\vec{S}$ (classically, $\vec{m}$) does
not simply rotate as $H$ is ramped.
The second possibility 
would mean that the spin
character of the many-body 
\linebreak
\begin{figure}
\vspace{-1.0 cm}
\begin{center}
\leavevmode
\epsfxsize=8.5 cm
\epsfbox{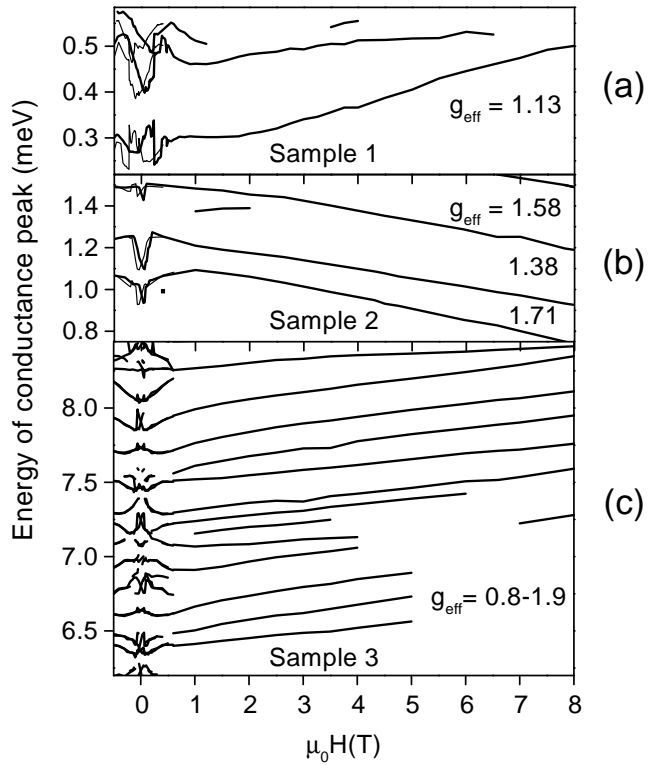}
\end{center}
\vspace{-.9 cm}
\caption{
\label{figure4}
Tunneling energies over a larger range of $H$.}
\end{figure}
\noindent
electron ground states
could evolve with $H$, so it
is possible that the threshold events for tunneling may consist of
$|\!\langle$$S_H\rangle\!|$-increasing transitions for some 
values of $H$, 
and $|\!\langle$$S_H\rangle\!|$-decreasing
transitions elsewhere.

For $\mu_0 H\!>\!2$ T, all measurable transition energies within a given
sample have the same sign of shift as a function  of $H$. This is
different from results in Al and Au nanoparticles, for which Zeeman spin
splitting of each orbital state gives rise to both upward and
downward-trending states vs.~$H$, with comparable
conductance amplitudes, and  with a degeneracy at $H\!=\!0$.  The
absence of spin degeneracy is not surprising in Co, since the exchange field
breaks the symmetry between spin-up and down.  However, since Co is not
fully spin-polarized ($P \approx 30\%$, \cite{Meservey}), the absence of
observed peaks shifting in both directions is an indication that electrons do not
couple equally well to all allowed many-body energy levels.  One factor may be that
Clebsch-Gordan coefficients will be very different for
$|\!\langle$$S_H\rangle\!|$-increasing
and $|\!\langle$$S_H\rangle\!|$-decreasing transitions in a system with a large
ground-state
value of the total electron spin, so that some tunneling matrix elements
may be immeasurably weak.  
Note that there is a sign difference between the large-$H$ slopes for sample 2 
and those of samples 1 and 3.  This might reflect a difference between electron-addition and electron-removal tunneling processes, a possibility which could be tested by fabrication of gated devices.

We wish to distinguish the linear dependence on $H$ that we measure for the
tunneling resonances at large $H$ from the linear
shift in the chemical potential
of micron-scale Co islands described by Ono {\em et al.} \cite{Ono}.  The
data of \cite{Ono} can be understood in terms of level crossings in a
magnetic island
with a continuum of energy
states.  As $H$ is ramped, spin-up levels cross with spin-down on
account of the Zeeman energy, and because the densities of the spin levels
are different, more states move one way than the other and the
chemical potential must shift.
This model is not applicable in
explaining the shifts of the {\em individual} levels that we observe,
because in our case any level crossings should be
individually resolvable.
Quantitatively, the shifts examined by Ono {\em et al}.~in Co correspond to
an effective g-factor of 0.7 \cite{Ono}, smaller than the values we measure.
(See Fig.~4.)

Finally, we turn to the measured density of tunneling resonances shown in
Fig.~2.  For the range of particle sizes imaged by STEM, 1-4~nm in
diameter, the average level spacing predicted for simple
non-interacting electrons should be between 0.75 and 40~meV, given the
calculated density of states (including both sp and d bands) in Co 
of 0.88~eV$^{-1}{\rm
atom}^{-1}$\cite{papa}.  In our measured spectra, the energy
spacing between tunneling peaks is less than 0.2 meV.  Enhanced densities
of tunneling resonances have previously been seen in Al nanoparticles for
values of voltage much greater than the single-electron level
spacing, and they were explained as an effect of non-equilibrium
electron-hole excitations within the particle \cite{Agam}.  The data in
Fig.~2 for samples 1 and 2 are different from the Al
results, however, in that an
increased density of levels is observed for energies even below the
expected single-electron spacing, where electron-hole excitations should
not be produced during tunneling.  As explanation, we note that electron
excitations within a Co nanoparticle may include low-energy spin waves in
addition to the independent-electron-type excitations seen in Al.
Inelastic emission of spin waves during tunneling may directly contribute
new tunneling peaks, and/or non-equilibrium spin excitations generated
during tunneling may produce extra tunneling peaks by shifting
single-electron states \cite{Agam}.  As a check, we can estimate the minimum energy needed to
excite spin-wave modes.  For a spatially uniform mode ($k$=0), the
excitation energy can be calculated using the anisotropy term in Eq.~(1),
to be $\approx\!2K_{m}\mu_B$.  If we use the size of the jump in the
tunneling energy at $H_{\rm{sw}}$, $\Delta$$E \approx 0.05$~meV 
(Fig.~3) to estimate
$K_{m}$
(using $\Delta$$E \approx \mu_BK_m$, see Fig.~3(b,c)), 
we arrive at a value of 0.1 meV for the spin-wave energy, which
would explain the enhanced density of tunneling states.  The contribution
of exchange energy to the lowest-energy {\em non-uniform} spin-wave modes
can be estimated by quantizing the spin-wave dispersion curve of Co within
the size of a nanoparticle.  This gives an energy  $(300$~meV$) (a/d)^2$
where $a$ is a lattice spacing and $d$ is the particle diameter \cite{AM},
or $\approx$ 1~meV for a 4-nm particle.

	In conclusion, we have measured discrete tunneling
resonances in nm-scale ferromagnetic Co particles.  Magnetic
anisotropy causes each resonance energy to shift reproducibly by on the
order of 0.1~meV as $H$ is swept about the hysteresis loop.  This effect
may provide a means to probe the dynamics of magnetization reversal
in nm-scale particles, complementary to magnetic force microscopy
\cite{Lederman}, Hall magnetometry \cite{Lok}, and SQUID techniques \cite{Wernsdorfer}.  Qualitative
features of these shifts can be described by a simple model.
However, a full explanation of the measurements will require a more detailed understanding of the electronic states inside a ferromagnet, including at least the contributions of low-energy collective spin-wave excitations to the electron states, and  effects of Clebsch-Gordan coefficients in tunneling.

	We thank C.~T.~Black, F. Braun, R. A. Buhrman, M. H. Devoret, H.
Hurdequint, J.~A.~ Katine, A. Pasupathy, D. Salinas, J. Silcox, Y. Suzuki,
M. Thomas, S. Upadhyay, and J. von Delft.  Support: ONR N00014-97-1-0745,
NSF DMR-9705059, Packard Foundation, and the Cornell
Nanofabrication Facility.

\end{multicols}


\begin{thebibliography}{99}
\bibitem{many} At large voltages, more than one nanoparticle
contributes to
the current in each sample, so that the Coulomb staircase curves cannot
be
used to estimate capacitances.  At the low voltages shown in Fig.~2, only
one particle contributes because of Coulomb blockade in the others.
\end{thebibliography}
\end{document}